\documentclass[sigconf]{acmart}
\usepackage{graphicx} 
\usepackage{tikz}
\usetikzlibrary{shapes.geometric, arrows}
\usepackage{wrapfig}
\usepackage{geometry}
\usepackage{listings}
\usepackage{tcolorbox}
\usepackage{multirow}
\usepackage{fancyhdr}
\usepackage{enumitem}

\newcommand{\tableSim}{
\begin{table}[htb!]
\scriptsize
\begin{tabular}{|l|c|c|c|c|c|c|c|c|c|c|}
\hline
\textbf{File}                                                                                 & $\alpha_1$    & $\alpha_2$    & $\alpha_3$    & $\alpha_4$  & $\alpha_5$    & $\alpha_6$    & $\alpha_7$    & $\alpha_8$    & $\alpha_9$    & $\alpha_{10}$   \\ \hline
\textbf{LOC}                                                                                  & 44   & 69   & 47   & 62 & 45   & 44   & 46   & 43   & 43   & 83   \\ \hline
\textbf{\begin{tabular}[c]{@{}l@{}}SR\\ (all)\end{tabular}}                 & 32\%   & 40\%   & 40\%   & 0\%  & 26\%   & 60\%   & 75\%   & 80\%   & 60\%   & 12\%   \\ \hline
\textbf{\begin{tabular}[c]{@{}l@{}}SR\\ (subset)\end{tabular}}                      & 32\%   & 10\%   & 48\%   & 0\%  & 87\%   & 40\%   & 75\%   & 80\%   & 80\%   & 48\%   \\ \hline
\textbf{\begin{tabular}[c]{@{}l@{}}$\Delta$ sim. \\ ($\times 10^{-2}$)\end{tabular}} & 0.09 & 0.07 & 9.56 & 0  & 7.83 & 2.99 & 6.37 & 8.19 & 1.05 & 2.76 \\ \hline
\textbf{Result}                                                                               &      &   \textcolor{red}{$\downarrow$}   & \textcolor{green}{$\uparrow$}     &    &  \textcolor{green}{$\uparrow$}    &  \textcolor{red}{$\downarrow$}    &      &      &    \textcolor{green}{$\uparrow$}  &     \textcolor{green}{$\uparrow$} \\ \hline
\end{tabular}
\caption{Results from generated files, where SR combines code preservation and passed test cases, and $\Delta$ sim. represents the similarity difference between the selected subset and all samples.}
\label{tab:sim}
\end{table}
}

\AtBeginDocument{%
  }

\usepackage{titlesec}
\titlespacing{\section}{0pt}{3pt}{1pt}  
\titlespacing{\subsection}{0pt}{1.5pt}{1pt}  

\setcopyright{acmlicensed}
\copyrightyear{2025}
\acmYear{2025}

\begin{document}

\title{Leveraging LLMs for Automated Translation of Legacy Code: A Case Study on PL/SQL to Java Transformation}

\fancyhead[L]{Leveraging LLMs for Automated Translation of Legacy Code}
\author{Lola Solovyeva}
\email{o.solovyeva@utwente.nl}
\affiliation{%
  \institution{University of Twente}
  \city{Enschede}
  \country{The Netherlands}
}

\author{Eduardo Carneiro Oliveira}
\email{e.carneirodeoliveira@uu.nl}
\affiliation{%
  \institution{Utrecht University}
  \city{Utrecht}
  \country{The Netherlands}}

\author{Shiyu Fan}
\email{sfan289@aucklanduni.ac.nz}
\affiliation{%
  \institution{Eindhoven University of Technology}
  \city{Eindhoven}
  \country{The Netherlands}
}

\author{Alper Tuncay}
\email{alpertuncay70@gmail.com}
\affiliation{%
 \institution{Leiden University}
 \city{Leiden}
 \country{The Netherlands}}

\author{Shamil Gareev}
\email{shamil.gareev27@gmail.com}
\affiliation{%
  \institution{Eötvös Loránd University}
  \city{Budapest}
  \country{Hungary}}

\author{Andrea Capiluppi}
\email{a.capiluppi@rug.nl}
\affiliation{%
  \institution{University of Groningen}
  \city{Groningen}
  \country{The Netherlands}}


\renewcommand{\shortauthors}{Solovyeva et al.}

\begin{abstract}
The VT legacy system, comprising approximately 2.5 million lines of PL/SQL code, lacks consistent documentation and automated tests, posing significant challenges for refactoring and modernisation. This study investigates the feasibility of leveraging large language models (LLMs) to assist in translating PL/SQL code into Java for the modernised "VTF3" system. By leveraging a dataset comprising 10 PL/SQL-to-Java code pairs and 15 Java classes, which collectively established a domain model for the translated files, multiple LLMs were evaluated. Furthermore, we propose a customized prompting strategy that integrates chain-of-guidance reasoning with $n$-shot prompting. Our findings indicate that this methodology effectively guides LLMs in generating syntactically accurate translations while also achieving functional correctness. However, the findings are limited by the small sample size of available code files and the restricted access to test cases used for validating the correctness of the generated code. Nevertheless, these findings lay the groundwork for scalable, automated solutions in modernising large legacy systems.
\end{abstract}



\maketitle


\section{Introduction}
Modernization of legacy code is still one of the critical challenges in software engineering of modern world, especially for large undocumented systems~\cite{modernizingLegacyCode}. Many critical systems across various industries continue to rely on legacy code, particularly in languages like COBOL, Fortran, PL/SQL, and Assembly~\cite{Benjamin2022Automated}. A substantial portion of legacy systems within the U.S. federal government has been in operation for 30 to over 60 years, presenting considerable challenges in terms of efficiency, maintenance, workforce availability, and security~\cite{Office_2019}.  
\par One of the key challenges of code modernization is the scarcity of developers proficient in legacy languages: the cost of acquiring such expertise continues to rise due to the decreasing number of professionals specializing in legacy systems~\cite{Charette2016Dragging}. Moreover, manual translation is a labor-intensive process that increases operational costs and prolongs the reliance on legacy system experts. For example, the Commonwealth Bank of Australia invested five years and approximately \$750 million to transition its platform from COBOL to Java~\cite{Jiaqi2024hmCodeTrans}.
\par Beyond purely manual code translation, the rapid expansion of code data has driven advancements in machine learning-based program translation, offering a more cost-effective and efficient alternative. The rise of LLMs for code synthesis and translation has sparked optimism about AI's ability to enhance efficiency and reduce risks in software modernization~\cite{Jiaqi2024hmCodeTrans, undefined2017Modernization, Benjamin2022Automated}. However, developers remain cautious, especially for safety-critical legacy systems, where unintended bugs could pose significant risks~\cite{10.1145/3613904.3642596}. Although LLM-driven translation is not entirely error-free, the accurately generated code segments can serve as a valuable foundation for developers, significantly streamlining the translation process. By incorporating manual adjustments where necessary, these AI-generated outputs can reduce the cognitive and technical workload associated with code translation. This enhances overall efficiency, minimizes human effort, and decreases both operational costs and time investment, making the translation workflow more scalable and accessible.
\par The modernization challenges are exemplified by the VT legacy system, from a large Dutch financial institution, which contains about 2.5 million lines of PL/SQL code, lacking consistent documentation and automated tests. In this state, the codebase needs to be modernized to Java for the "VTF3" system to maintain a quality in an evolving technological landscape. This paper aims to evaluate the potential of LLMs for automating the translation of such massive legacy systems and consequently reduce the human effort involved in modernizing legacy codebases. Our goal is therefore focused on establishing a methodology resulting in a codebase that is syntactically correct and functional within the requirements of the newly modernized code. 
The contributions of our study are threefold:
\begin{itemize}[topsep=2pt, partopsep=0pt, itemsep=2pt, parsep=0pt]
    \item An evaluation of the ability of LLMs to translate PL/SQL code into Java using prompting strategies.
    \item The development of a prompting methodology aimed at improving both the syntactical and functional accuracy of the translated code.
    \item An analysis of key factors that influence the successful translation of code.
\end{itemize} 
\par Our findings indicate that incorporating a domain model, which guides the LLM to generate code in a specific format, and using a targeted set of translation examples, leads to more effective outcomes. Additionally, the success of our approach depends more on the similarity of the provided samples to the translation task than on their quantity. Finally, this method is adaptable to other language-to-language translation tasks beyond PL/SQL and Java, as it is not language-specific.


\begin{figure*}[t]
    \centering

\begin{tikzpicture}[node distance=3cm]

\tikzstyle{startstop} = [rectangle, rounded corners, minimum width=2cm, text width = 2.5 cm, minimum height=1.5cm,text centered, draw=black, fill=red!30]
\tikzstyle{process} = [rectangle, minimum width=2cm, text width = 2.5 cm,  minimum height=1.5cm, text centered, draw=black, fill=orange!30]
\tikzstyle{arrow} = [thick,->,>=stealth]

\node (start) [startstop] {Provide context to the LLM};
\node (step1) [startstop, right of=start] {Ask for a prompting strategy};
\node (step2) [process, right of=step1] {Provide domain model};
\node (step3) [process, right of=step2] {Provide PL/SQL-Java pairs};
\node (step4) [process, right of=step3] {Ask for a translation};

\draw [arrow] (step1) -- (step2);
\draw [arrow] (step2) -- (step3);
\draw [arrow] (step3) -- (step4);

\draw [arrow, bend left=30] (start) to (step1);
\draw [arrow, bend right=-30] (step1) to (start);

\end{tikzpicture}

    \caption{A flowchart illustrating the pipeline of interaction with the LLMs through prompting and chain-of-guidance.}
    \label{fig:methodology}
\end{figure*}
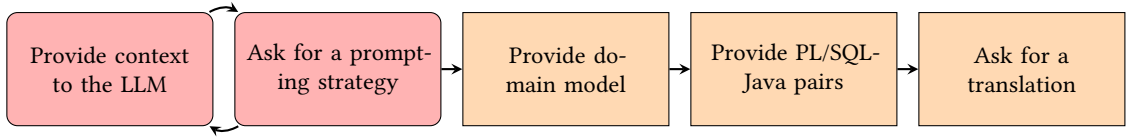
\section{Related Work}
\label{sec:_related}
The rapid advancement in LLMs has significantly impacted various domains of software engineering, particularly in automating complex tasks like code translation, modernisation and maintenance. This section reviews key contributions in three areas: the role of LLMs in software maintenance and legacy systems; recent approaches to automating code translation and modernisation; and the evaluation and benchmarking of LLMs for code transformation tasks.

Recent research highlights the integration of LLMs in software engineering for tasks like requirements engineering, coding assistance, testing and maintenance ~\cite{Gao2025Current, He2025LLM, Wang2024Software}. LLMs can enhance productivity by automating tasks such as code generation, code completion and comment classification~\cite{Wong2023Natural,Patel2025State}. However, some challenges remain regarding model robustness, ethical considerations and the need for human oversight~\cite{Marques2024Using,France2024Navigating}. Frameworks for assessing LLM `readiness'~\cite{Patel2025State} and multi-agent systems for addressing complex software engineering challenges~\cite{He2025LLM} have been proposed.  Our study complements this body of work by demonstrating a practical implementation of LLM-driven code modernization, evaluating its feasibility in a real-world large-scale legacy system.

Code translation and modernization are crucial for updating legacy systems and improving software efficiency. Various approaches have been proposed, including human-machine collaboration for interactive code translation~\cite{Jiaqi2024hmCodeTrans} and incremental modernization of legacy CFD codes~\cite{undefined2017Modernization}. Neural-guided program synthesis has shown promise in automating the transpilation of imperative to functional code~\cite{Benjamin2022Automated}. Microservices architecture has been explored as a strategy for legacy software modernization, offering improved maintainability and scalability~\cite{Knoche2018Using}. 
We advance this work by using LLMs to automate PL/SQL-to-Java translation, refining prompting techniques and introducing cosine similarity-based methods to improve translation accuracy.

The evaluation of LLMs for code generation and summarisation is an emerging field.
Several studies propose frameworks and methodologies for assessing LLMs' code synthesis capabilities, including functional and non-functional qualities ~\cite{Sagodi2024Methodology,Yeo2024Framework}. Researchers have also developed benchmarks for evaluating LLMs in specific domains, such as programming service mobile robots ~\cite{Hu2024Deploying} and bug reproduction~\cite{kang2024evaluating}. The importance of comprehensive evaluation across multiple dimensions, including validity, legality, and readability, has been emphasized~\cite{Chen2025VisEval}. Studies have highlighted the significance of instruction tuning and the need for high-quality reference data in summarization tasks~\cite{Zhang2024Benchmarking}. Additionally, researchers have called for multi-prompt evaluations to address the brittleness of single-prompt assessments~\cite{Mizrahi2024State}. 
Our work contributes to this debate by evaluating LLMs for PL/SQL-to-Java translation, and by introducing a structured pipeline for optimising translation quality using similarity metrics and domain-specific context selection.

\section{Methodology}
\label{sec:methodology}


\par \textbf{Dataset.} The dataset consisted of ten pairs of code examples, each containing a PL/SQL version and its corresponding Java equivalent. Additionally, the domain model, implemented in Java as part of the VTF3 framework, comprises ten classes, three interfaces, and two records. All components within the domain model are related to the target files that require translation into Java. This domain model serves as a constraint, guiding the language model to generate files that conform to the expected structural and semantic specifications. 
\par \textbf{Setup.} We developed an initial prompting strategy and explored the code translation capabilities of five LLMs - GPT 4o, o1, Claude 3.5 Sonnet, Tabnine, and DeepSeek-R1.  Considering its potential compared to other techniques \cite{brown2020language}, we adopted the few-shot prompting technique in the initial stage. Due to the context window limitations, we could only provide two pair samples to the LLM. An example of a prompt is shown below: 
\definecolor{background}{rgb}{1,0.98,0.9}
\definecolor{border}{rgb}{0.3,0.3,0.3}
\definecolor{codebg}{rgb}{0.95,0.95,0.95}
\begin{tcolorbox}[
    colback=background,
    colframe=border,
    arc=2mm,
    coltitle=black
]
\textbf{Human:} 
Given the 2 examples of translation of PL/SQL to Java: \texttt{1.plsql} \texttt{1.java} and \texttt{2.plsql} \texttt{2.java}. Give me the Java translation of the following PL/SQL file: \texttt{query.plsql}\\
\textbf{Assistant:}
\texttt{response.java}
\end{tcolorbox}

\par \textbf{Prompt refinement.} Based on the preliminary findings from our experiments with two-shot prompting, we refine our prompting strategy to enhance performance. Given that the initial results highlighted the superiority of DeepSeek-R1, we selected this model for further prompt optimization, implementing a structured pipeline. In our developed pipeline, we give the problem context - i.e. the task and the available dataset - to the LLM and explicitly ask for the best prompting strategy. We refine the context and again ask for a prompting strategy. This cycle repeats until the LLM suggests an adequate prompting strategy for code translation based on the specific problem context. Figure \ref{fig:methodology} illustrates the pipeline developed in this study.

\par We adopted the prompting strategy recommended by the LLM which is divided into three parts: 1) it begins with providing the Java classes related to the domain model; 2) followed by presenting pairs of code translation samples from PL/SQL to Java; and 3) concluding with querying the model for one specific PL/SQL file to be translated to Java. This approach incorporates both chain-of-guidance \cite{rajimproving} and $n$-shot prompting techniques. Regarding the number of samples included in the prompt, we performed a comparative test between providing all available samples versus selecting only those samples that are most syntactically similar to the file to be translated based on cosine similarity.
\par \textbf{File similarity analysis.} To enhance code translation accuracy, we hypothesized that providing multiple PL/SQL examples as input to the LLM would yield a more accurate Java translation. Our initial approach revealed that inputing two or nine code examples did not consistently result in the highest translation accuracy. We hypothesized that higher translation accuracy is expected when the input code exhibits greater similarity to the target output code.
\par To quantify this relationship, we used cosine similarity~\cite{XIA201539} to measure the similarity between the combined input PL/SQL code examples and the corresponding PL/SQL code of the generated Java output. We systematically evaluated all 502 possible combinations of the 10 PL/SQL code examples to identify the combination that achieved the highest similarity score. This optimal combination was then used as input to translate into Java code, aiming to maximize translation accuracy.
\par
\begin{figure}[t!]
    \centering
    \includegraphics[width=\linewidth]{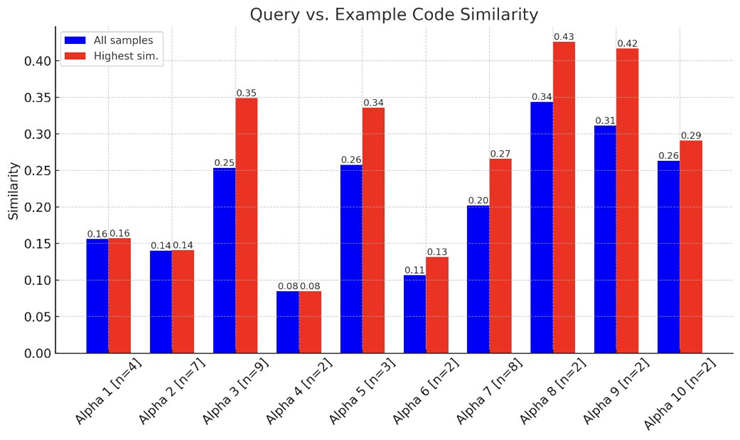}
    \caption{Highest similarity values (red) vs. using all examples (blue) for PL/SQL generation}
    \label{fig:highest-similarity}
\end{figure}

 Figure~\ref{fig:highest-similarity} illustrates the similarity values of the example files relative to the query file. The similarity scores for all the provided examples are represented by blue bars, while the combination yielding the highest similarity score is highlighted in red. 

To evaluate our hypothesis—that higher similarity between sample files and the query file enhances the quality of the generated output—we conducted two experimental runs for each generated file: one using all nine sample files and the other using only the samples that achieved the highest similarity scores.
\par \textbf{Evaluation.} To assess the accuracy of the generated files, prewritten test cases developed by the company responsible for VT were used to evaluate whether their functionality aligned with the expected outcomes. 
Additionally, due to an initial data anonymization, the generated files underwent a de-anonymization process, which required manual adjustments to certain parts of the file. Also, the files were assessed based on the extent to which the generated content required modification before successful integration into the system. Therefore, we employed two metrics to evaluate the quality of the generated files: the percentage of code preserved ($CP$) and the percentage of test cases passed ($TP$). These metrics were integrated into a unified measure, referred to as the Success Rate ($SR$), by computing the product of their respective values as shown in the following equation.
\begin{equation}
    SR(\%)=\frac{CP \times TP}{100}
\end{equation}
\section{Results}
\label{sec:results}
\begin{figure}[htb!]
    \centering
    \includegraphics[width=\linewidth]{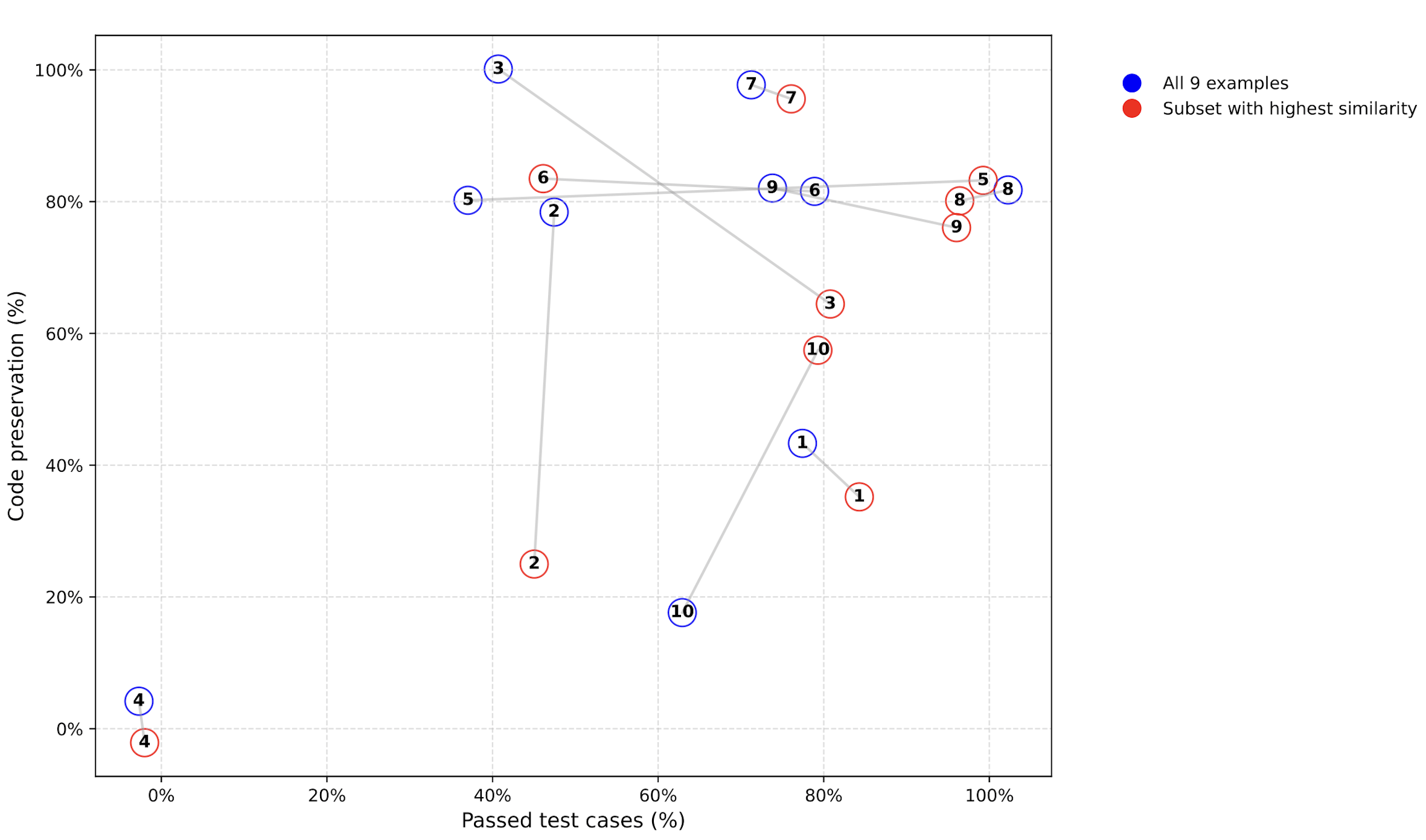}
    \caption{Results for code preservation and percentage of passed tests cases for generated files}
    \label{fig:results}
\end{figure}

Figure \ref{fig:results} compares the preservation of code and the test cases passed on ten generated files. Blue represents results from generated files using 9-shot prompting, while the red color indicates that only a subset of samples—those most similar to the query file—was used. The majority of red circles are positioned toward the right, indicating a higher number of passed test cases compared to blue marks. In some cases, the blue circles are positioned above the red, indicating a higher degree of code preservation compared to the red samples.

\par A key observation from the results is that providing the maximum number of available examples leads to high code preservation. However, the functional correctness of the generated code remains relatively low. In cases where functional correctness is the primary objective, selecting only the samples that are most similar to the query file proves to be the most effective approach. 
\tableSim
\par Table \ref{tab:sim} summarizes the link between sample similarity and successful execution. The results indicate that in four out of ten cases, the quality of the generated file improved, demonstrating higher syntactic accuracy and functionality aligned with expectations. In the two cases where quality decreased, specifically for $\alpha_2$ and $\alpha_6$,  low initial similarity led to quality declines, as observed in Figure \ref{fig:highest-similarity}. A similar pattern is observed for $\alpha_4$, where it showed no improvement due to identical similarity values. In the cases of $\alpha_7$ and $\alpha_8$, the generated file already exhibited satisfactory performance with all samples, and thus, no significant enhancements were observed for a subset.
\par \textbf{Findings.} The study demonstrated that LLMs can support the translation of PL/SQL code into Java, with varying degrees of success. First, it becomes evident that incorporating the domain model—which constrains the LLM to generate files in a specific format—along with providing a defined set of translation examples, leads to a more effective and successful strategy. Secondly, our results suggest that the effectiveness of the approach is determined not by the number of samples provided but by their similarity to the sample that requires translation. Finally, this approach to language-to-language translation is not limited to PL/SQL and Java, as it does not rely on language-specific components. Therefore, the same pipeline can be adapted for other language-to-language translation tasks. 
\par \textbf{Implications for Practitioners.}
The adoption of LLM-based methodologies for legacy code modernization presents unique opportunities and challenges: insights from an interview with the company responsible for the VT system highlighted several critical implications.

The primary motivation for using LLM-based transpilation over manual or rule-based approaches lies in its scalability. Translating 2.5 million lines of undocumented PL/SQL code manually would be prohibitively time-intensive and costly. LLMs offer the flexibility to bridge paradigm differences between languages like PL/SQL and Java, while restricting the problem space through structured interfaces and APIs mitigates hallucination risks, enabling scalable automation.

Although the project remains conceptual, it reflects a commitment to achieving efficiency gains. Hiring interns to refine the techniques shows the expectation of long-term returns, though concrete metrics are yet to be established. Challenges such as integration with legacy systems, model unpredictability, and privacy concerns remain unresolved but are anticipated as pivotal.

Maintainability of translated Java code is expected to be a major challenge. Strategies such as static code analysis, rigorous linting, and consistent human oversight are critical to ensuring quality. Carefully crafted LLM prompts that enforce coding style will also support sustainable outcomes.


For the case owner and other practitioners, LLM-based modernization holds promise for addressing scalability, cost, and efficiency challenges in legacy systems, but its success depends on overcoming significant technical and organizational hurdles.
\par \textbf{Limitations.} One limitation of this study is the small sample size, which affected the scalability of our approach. Also, the non-deterministic behaviour of LLMs may impact reproducibility. Finally, we used text-based metrics to evaluate code similarity. For studies with more comprehensive datasets, we recommend using other evaluation metrics, such as abstract syntax trees.

\section{Conclusion}
\label{sec:conclusion}
This study has demonstrated the potential of using LLMs for automating the translation of PL/SQL legacy code to Java, particularly through the use of task-specific prompting, and similarity-based input selection. The approach shown produced in majority cases both syntactically correct, and functionally aligned translations: however, there were challenges such as the limitation of our dataset, the non-deterministic behaviour of the chosen models and their scalability.

As future work, we plan to expand the dataset to improve the generalisability of our approach, and to integrate more evaluation metrics. We also plan to combine LLMs in multi-modal strategies, to enhance the accuracy and the reliability of the translations.

\section*{Acknowledgements}
The authors acknowledge the use of AI-based tools to enhance the clarity and grammatical accuracy of sentences.

\bibliographystyle{plain}
\bibliography{main}
\end{document}